\documentclass[slac_one]{revtex4}
\usepackage{graphicx}
\usepackage{epsfig}
\usepackage{fancyhdr}
\begin{document}

\title{{\small{Hadron Collider Physics Symposium (HCP2008),
Galena, Illinois, USA}}\\ %% Please keep this conference title here
\vspace{12pt}
Top quark in theory} %% Paper title goes here


\author{Eric Laenen}
\affiliation{Nikhef, Theory Group,
Kruislaan 409, 1098 SJ Amsterdam, The Netherlands\\
University of Amsterdam, ITFA ,
Valckeniersstraat 65, 1018 XE Amsterdam, The Netherlands\\
University of Utrecht, ITF
Leuvenlaan 4, 3584 CE Utrecht, The Netherlands}

\begin{abstract}
I review how the top quark is embedded in the Standard Model and 
some its proposed extensions, and how it manifests itself in various
hadron collider signals.
\end{abstract}

\maketitle

\thispagestyle{fancy}


\section{TOP IS SPECIAL} % Section title should be in all capitals.

Of the particles seen so far in collider experiments, the top quark is no
doubt the most expensive, and the most glamorous. It is therefore the
center of attention at the Tevatron and the LHC, until of course a new
star, Higgs, comes along.  
Expressed less colloquially, the top quark is considered an interesting study object 
because it has many quantum numbers and thus couples to almost all other particles,
through various (chiral, vector, scalar) structures, all of which bear scrutiny
for deviations.  Precise scrutiny is feasible because the large top mass
implies, first, that it couples strongly to whatever breaks the electroweak
symmetry, and second, the resulting large width minimizes obscuring
hadronization effects and allows preservation of spin information.
Top is also a troublemaker for the Standard Model, contributing
significantly to the quadratic divergences of the Higgs self energy,
but is at the same time an life raft for beyond the Standard Model (BSM)
theories such as the MSSM (raising the upper limit on the light
Higgs in that theory).
With the Tevatron having made the first precious thousands top quarks, 
leading to its discovery and ftests of some of its properties,
the LHC is a genuine top quark factory and will allow us to study the top quark in great detail.  
Here I review some of the interesting aspects of top quark physics. I will
necessarily be short on length and details, and I refer to 
other excellent recent reviews \cite{Bernreuther:2008ju,Han:2008xb,Quadt:2007jk,Chakraborty:2003iw,Beneke:2000hk} for more.

\section{TOP QUARK IN THE STANDARD MODEL AND BEYOND}
\label{sec:top-quark-standard}

\subsection{Standard Model}

Let us recall the various interactions of the top quark field $t(x)$ 
 in the Standard Model Lagrangian.
The interaction with gluons is a vectorlike coupling involving
an $SU(3)$ generator in the fundamental representation
\begin{equation}
  \label{eq:1}  
g_s \bar{t}_i(x) \gamma^\mu \left[T^a\right]_{ij} t_j(x) G_\mu^a(x)\,,
\end{equation}
where $i,j$ label color. The interaction with photons is simply vectorlike and proportional
to the top quarks fractional charge
\begin{equation}
  \label{eq:2}
  \frac{2}{3}e\,\, \bar{t}(x) \gamma^\mu   t(x) A_\mu(x)\,.
\end{equation}
Its charged weak interaction is chiral and flavor-changing
\begin{equation}
  \label{eq:3}
  \frac{g_w}{2\sqrt{2}} V_{ti}\,\, \bar{t}(x) \gamma^\mu(1-\gamma_5) i(x) W_\mu(x),
\quad i={d,s,b}\,,
\end{equation}
while it neutral weak interaction is flavor-conserving  and
parity violating
\begin{equation}
  \label{eq:16}
  \frac{g_w}{4\cos\theta_W} , \bar{t}(x) \gamma^\mu\left((1-\frac{8}{3}\sin^2\theta_W) -\gamma_5\right) t(x) Z_\mu(x)\,.
\end{equation}
The interaction of the top with the Higgs boson of the Yukawa type
\begin{equation}
  \label{eq:4}
  y_t\,\, h(x) \bar{t}(x) t(x)
\end{equation}
with a coupling constant that is directly related to its mass
$y_t = \sqrt{2}m_t/v$.

Beyond these, effective interactions such as for flavor-changing neutral 
currents, occur due to loop corrections, and are therefore very small.
All of these couplings could be modified in structure and strength 
by virtual effects due to new interactions associated with physics 
beyond the Standard Model. This is particularly relevant for the top
quark if only because it evidently couples strongly to the electroweak
symmetry breaking sector (the Yukawa coupling $y_t$ in Eq.~(\ref{eq:4}) is
very close to 1 in strength).  It is therefore important to test
these structures in detail. Such studies can be guided by educated guesses
about possible alternatives to the Standard Model, 
and so let us briefly review the role of the
top quark or its partners in some BSM models.

\subsection{Beyond}
\label{sec:beyond}

\subsubsection{Supersymmetry}
\label{sec:supersymmetry}

If it weren't for the top quark corrections to the lightest Higgs boson
mass, the Minimal Supersymmetric
Standard Model (MSSM) would predict it to be lighter than the $Z$ boson,
and would thus already have been ruled out. The maximum viable mass
for this boson is thus about $140$ GeV. 
Top plays an even more central role, in that its dominant contribution
to the running of a Higgs potential parameter down from the GUT scale in fact
leads to a negative eigenvalue for the Higgs mass matrix, thereby
explaining electroweak symmetry breaking. Moreover,
in the supersymmetry searches at the LHC, both regarding
discovery and subsequent unraveling,  top would 
play a key part, as many heavy supersymmetric partners have
top among its decay products.

\subsubsection{Little Higgs}
\label{sec:little-higgs}

The key idea \cite{ArkaniHamed:2001nc,ArkaniHamed:2002pa}
is to construct a model in which the Higgs boson emerges
as a pseudo-Goldstone boson, and is therefore 
naturally light. This is analogous to explaining 
the lightness of the pion by its nature as a pseudo-Goldstone boson
for spontaneously broken chiral symmetry. A number of such Little Higgs models have in
fact now been constructed, see \cite{Schmaltz:2005ky} for a review. In these
models the top quark plays a key role, simply because its contribution to the Higgs
mass corrections are the most dominant, and must be cancelled
to an appropiate extent, i.e. up to a certain scale. In fact, ambitions here do not run
as high as the GUT scale; the goal is to solve at least the ``little hierarchy''
problem, keeping the Higgs mass natural up to a scale a factor 10 above
what would still be natural for the Standard Model, up to about 10 TeV.
Perhaps the most extensively explored model is the Littlest Higgs model with $T$-parity
\cite{Cheng:2003ju,Cheng:2004yc}. Here, heavier mirror copies of the four Standard Model electroweak
gauge fields appear, which cancel the Higgs mass contributions of the latter.
Likewise, T-odd and T-even fermionic top-partners are introduced whose 
Higgs mass contribution cancel, to one-loop, that of the top. The heavy top
partners, that also decay to tops, could be visible by LHC experiments.

\subsubsection{Extra dimensions}
\label{sec:extra-dimensions}

In extra-dimensional scenarios \cite{ArkaniHamed:1998rs,Randall:1999ee}
top plays less of a central role, as the mechanism
for curing the hierarchy problem is not based on particles but spacetime. Nevertheless,
the Kaluza-Klein excited states of gluons might be best visible 
as resonances in top quark pair production channels, as these are more easily
identified. 

\subsubsection{Top condensation}
\label{sec:topcolor}

Top has also played a key role in setting up models in which, 
in analogy with BCS superconductively, the Higgs is effectively
a fermion (top) boundstate, formed by new ``topcolor'' gauge interaction
\cite{Hill:1991at,Hill:1993hs}
that views  the 3rd generation as special. Evidence for such a mechanism,
which would also yield charged and neutral top-pions, and possibly
new heavy gauge bosons, could arise in $t\bar{t}$ invariant mass
distributions. By their nature, these new particles would couple strong
to top quarks, and lead to a varied phenomenology involving
top final states. A comprehensive review is  Ref.~\cite{Hill:2002ap}.

\section{TOP MASS}
\label{sec:top-mass}

The top quark property that is most readily employed in top physics
is its mass. The Tevatron experiments have set the standard to a level
that will be hard to pass by the LHC by measuring it 
to less than 1\% accuracy ($172.6\pm 1.4$ GeV).
Together with an accurately measured $W$ boson mass it 
severely constrains the mass range of a possible Higgs boson both
in the Standard Model and in the MSSM. Therefore its precise measurement is
of considerable importance, and so also its careful definition.
A natural definition is based on the location of the pole of the full 
top quark propagator, the pole mass. However, because 
the top quark, being colored, can never propagate out to infinite times - 
a requirement for the definition of
a particle mass in scattering - such a pole only exists in perturbation
theory, and its location is intrinsically ambiguous by $\mathcal{O}(\Lambda_{QCD})$
\cite{Beneke:1994sw,Bigi:1994em,Smith:1997xz}. A theoretically more
precise definition is the $\overline{MS}$ mass $\bar{m}(\mu)$  whose
relation to the pole mass is known to sufficiently high order. For $\mu$
one often takes the implicit value found when intersecting the $\bar{m}(\mu)$
curve with the $\bar{m}(\mu)=\mu$ axis, yielding $\bar{m}(\bar{m})$.
For the top quark, this value is about 10 GeV smaller than the pole
mass, and thus the question often arises what mass the Tevatron and 
LHC experiments measure. Experimentally, the top quark mass is reconstructed 
by collecting jets and leptons. Soft particles arising from both within and 
outside these jets may enter them, and thus affect the reconstructed mass. 
Moreover, various experimental methods used (e.g. track quality cuts), and 
Monte Carlo based corrections, do not have a clean perturbation theory description. 
Therefore the question is difficult to answer, but the pole mass should be the closest to 
the true answer.

\section{TOP CROSS SECTION}
\label{sec:top-cross-section}

The top quark inclusive cross section at hadron colliders has been 
a constant of theoretical attention over many years, with steady progress
toward its more accurate determination. Let us look at some recent
developments, for which we need a brief discussion on threshold resummation.

\subsection{Threshold resummation}
\label{sec:thresh-resumm}

When the top quark pair is produced near threshold, logarithms
whose argument represents the distance to threshold in the perturbative
series become numerically large. It is important to note here that
the definition of the threshold depends on the observable. Thus, 
for the inclusive cross section threshold is given by
$T_1:\, s-4m^2=0$. For the transverse momentum distribution we have
$T_2:\, s-4(m^2+p_T^2) = 0 $, and for the doubly differential distribution in 
$p_T$ and rapidity we have $T_2:\, s-4(m^2+p_T^2)\cosh y=0 $. The perturbative
series for any of these (differential) cross sections can be expressed as
\begin{equation}
  \label{eq:7}
  d_\alpha \sigma (T_\alpha) = \sum_n \sum_k^{2n} \alpha_s^{n} c^\alpha_{n,k} \ln^k (T_\alpha)\,,
\end{equation}
plus non-logarithmic terms.
Here $T_\alpha$ represents any of the threshold conditions, suitably
normalized, for the observables
enumerated by $\alpha$. Note that it is of course allowed to use e.g. $T_2$ for the
inclusive cross section, by first analyzing $d\sigma/dp_T$ and then integrating
over $p_T$. For any complete fixed order calculation this will give the same 
answer, but if one only selects the logarithmic terms because the exact answer
is unknown, numerical differences will occur which can be seen as an
theoretical uncertainty \cite{Kidonakis:2001nj}.

The logarithms result from phase space regions where the extra gluons emitted
are soft and/or collinear to their massless, on-shell emitter.
Resummation concerns itself with carrying out the sum in Eq.~(\ref{eq:7}). 
To this end it is often convenient, in order to account for momentum or energy conservation
conditions, to take moments with respect to
$T_\alpha$
\begin{equation}
  \label{eq:8}
 d\sigma (N)  = \int dT_\alpha T_\alpha^N   = \sum_n \sum_k^{2n} \alpha_s^{n} \tilde{c}^\alpha_{n,k} \ln^k N
\end{equation}
The resummed result then takes the generic form
\begin{equation}
  \label{eq:9}
   d\sigma (N) = \exp \left(Lg_1(\alpha_sL) + g_2(\alpha_sL) + \alpha_sg_3(\alpha_sL)
+\ldots \right)\times C(\alpha_s) \,.
\end{equation}
After resummation, the inverse transform to Eq.~(\ref{eq:8}) should
be taken.  Including up to the function $g_i$ in the exponent amounts to N$^i$LL resummation.
Not surprisingly, for higher $i$ these functions are progressively more difficult to 
determine.
Key benefits of resummation are (i) gaining all-order control of the terms that
make using fixed-order perturbation theory unreliable, thereby restoring predictive
power, and (ii) reduction of scale uncertainty. For Drell-Yan or Higgs production via
gluon fusion, the threshold is $T: s-Q^2=0$, and the resummed partonic cross section
has the form \cite{Sterman:1986aj,Catani:1989ne,Catani:2003zt,Forte:2002ni,Eynck:2003fn,Laenen:2005uz,Becher:2007ty,Ravindran:2006cg,Idilbi:2006dg}
\begin{equation}
  \label{eq:10}
  d\sigma(N) = C(\alpha_s) \times
\exp\left[\int_0^1 \frac{z^{N-1}-1}{1-z}\left\{ 
2\int_{\mu_F^2}^{(1-z)^2Q^2} \frac{d\mu^2}{\mu^2}
A_i(\alpha_s(\mu^2)) + D_i(\alpha_s(1-z)^2Q^2)
\right\} \right]\,.
\end{equation}
The function $A_i$ is the cusp anomalous dimension and controls
soft and collinear radiation, while $D_i$ includes
contributions from soft, wide-angle emissions. Both are known to 3rd order.
The resummed top quark pair production cross sections with threshold $T_1$
is similar in form, but here $D$ is a matrix in color space.

Most of the latest theoretical estimates for the inclusive pair production cross sections
take threshold resummation to a certain accuracy into account. Furthermore, 
uncertainties due to scale variables (lessened by the resummation), and the
PDF's are included. It should be said however that, while the top is always pair-produced
fairly close to threshold at the Tevatron, at the LHC this is not necessarily true, and hence
the impact of threshold-resummation is smaller.

Thus, Cacciari et al. in \cite{Cacciari:2008zb} update their earlier \cite{Cacciari:2003fi}
estimates for the inclusive pair production cross section for a range
of masses, for number of PDF sets,and  for both NLO and NLO+NLL cross
sections. They find that at the LHC the scale
uncertainties are significantly larger than the PDF uncertainties, and
that it is important to vary the renormalization $\mu_R$ and
factorization scale $\mu_F$ independently.
Nadolsky et al in Ref.\cite{Nadolsky:2008zw}, in the context of using
the CTEQ6.6 set, examine the potential of the $t\bar{t}$ cross section 
as a gluon density probe, and its role in normalizing certain classes
of LHC cross sections.
Kidonakis and Vogt \cite{Kidonakis:2008mu} derive NNLO estimates for
the inclusive cross section from the NLL resummmation for the double differential cross section, 
and include uncertainties due to PDF, scale, and kinematics choice. 
Moch and Uwer \cite{Moch:2008qy} also update the NLO-NLL results of 
the results in Ref.~\cite{Cacciari:2003fi} and then extend the resummation
accuracy to NNLL, using threshold $T_1$. The required 3-loop $A_i$ 
\cite{Moch:2004pa,Vogt:2004mw} and
the 2-loop $D^{t\bar{t}}_i$ which they were able to constitute
from the Drell-Yan and Higgs equivalent functions
 \cite{Vogt:2000ci,Contopanagos:1997nh}, and results and insights from
Refs.~\cite{Mitov:2006xs,Bernreuther:2004ih,MertAybat:2006mz}. 
From this result they construct a second order estimate for the $t\bar{t}$
inclusive cross section of the form
\begin{equation}
  \label{eq:11}
  \sigma^{t\bar{t},(2)} = \alpha_s^2 \sum_{n=0}^4 c_n \ln^n \beta,
\quad \beta = \sqrt{1-\frac{4m^2}{s}} \; +\; \mathrm{Coulomb\;corrections}
\end{equation}
with $c_4,\ldots,c_0$ all known, including the Coulomb corrections.
Furthermore, all scale-dependent logartithms, constructed using
renormalization group methods, were included. A much reduced
scale uncertainty was found, albeit for $\mu_F$ and $\mu_R$ held equal.

Results such as those above serve to whet the appetite for an exact NNLO
calculation for the inclusive cross section, and also on this front much progress is being made. 
The real calculations are now available, thanks in large measure to
having the NLO $t\bar{t} + \, \mathrm{jet}$ available \cite{Dittmaier:2007wz}. 
The 2-loop virtual corrections are have been obtained in the limit
$s,|t|,|u| \gg m_t^2$ from both factorization-based calculations \cite{Mitov:2006xs},
and direct calculation using semi-automatized Mellin-Barnes 
techniques \cite{Czakon:2007wk}. More recently \cite{Czakon:2008zk}, the first full
2-loop virtual results have been derived. The exact NNLO calculation
seems to be no longer infinitely far on the horizon.

\section{SINGLE TOP}
\label{sec:single-top}

\begin{figure}
  \centerline{\includegraphics[width=0.55\columnwidth]{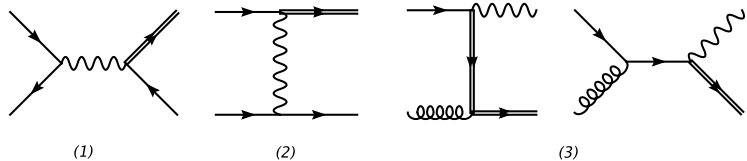}}
  \caption{From left to right the $s$-channel (1), $t$-channel (2) processes,
    and the $Wt$ associated (3) production
    channel.}\label{Fig:singletopprocesses}
\end{figure}

Single tops are produced by the weak interaction, in
processes that are customarily categorized (Fig.~\ref{Fig:singletopprocesses})
using Born kinematics.
A particularly interesting aspect of single-top production is the
prospect of directly measuring $V_{tb}$ and testing the chiral
structure of the associated vertex: top produced singly in this way
is highly polarized, and offers a chance to study its spin.
Furthermore, the dominant $t$ channel at the LHC will, when 
confronting measurements with a 5-flavor NLO calculation,
allow extraction of the $b$-quark density. This will be useful
in predicting other production processes at the LHC.
The single top production characteristics are sensitive to new physics,
depending on the channel. Thus, the $s$-channel will be sensitive to
e.g. $W'$ resonances, the $t$-channel to FCNC's. Experimentally, this
process turns out to be very difficult to extract from backgrounds, and so
far only (strong) evidence  has been found by the D0 \cite{Abazov:2006gd}
and CDF \cite{Aaltonen:2008sy}
collaborations, with 95\% CL lower limits on $V_{tb}$ of 
0.68 and 0.66, resp. The measured cross sections agree within
errors with the NLO calculations
\cite{Harris:2002md,Cao:2004ap,Cao:2005pq,Campbell:2004ch,Campbell:2005bb}.
The inclusive cross sections at the Tevatron are rather small,
0.9 ($s$) and 2 ($t$) pb, with the $Wt$ channel negligible. At the
LHC the numbers are, approximately, 10, 246 and 60 pb, respectively.
Clearly at the LHC, the $t$-channel will be dominant. Besides interesting
in its own right, this process is a background to putative new physics
processes, such as Higgs production in association with a $W$ boson.

\section{TOP DISTRIBUTIONS}
\label{sec:top-distributions}

While the inclusive cross section has received much theoretical and
experimental attention, the interest in distributions in certain variables
is increasing, given the increased Tevatron data set, and the LHC start.
Let us review some recent developments.

\subsection{Charge asymmetry}
\label{sec:charge-asymmetry}

The charge asymmetry is the difference in production rate for top and
anti-top at fixed angle or rapidity. While electroweak production via
a $Z$-boson could produce a (very small) asymmetry at LO, QCD itself
does produce it at $\mathcal{O}(\alpha_s^3)$ through a term proportional
to the SU(3) $d_{abc}$ 
symbol \cite{Nason:1989zy,Beenakker:1991ma,Kuhn:1998kw,Dittmaier:2007wz}
CDF and D0 have recently performed first measurements, 
albeit with still large uncertainties of this asymmetry
\cite{:2007qb,Aaltonen:2008hc}. Thus, the impact of even higher orders 
becomes interesting  which at this stage can only be assessed 
from approximate, resummation based calculations \cite{Almeida:2008ug,Kidonakis:2001nj} ,
and was studied in Ref.~\cite{Almeida:2008ug}.
The asymmetry was found to be stable with respect to including such higher order corrections, 
and to be much less sensitive to scale variations. At the LHC, where the $gg$ channel
dominates, the asymmetry is naturally small, but may be enhanced at large
invariant mass, where the $q\bar{q}$ channel regains strength. The charge 
asymmetry is present at leading order in $t\bar{t}+\,\mathrm{jet}$ production.
However, NLO corrections \cite{Dittmaier:2007wz} appear to wash out the
asymmetry for this reaction.

\subsection{Invariant mass}
\label{sec:invariant-mass}

\begin{figure}
  \centerline{\includegraphics[width=0.35\columnwidth,angle=90]{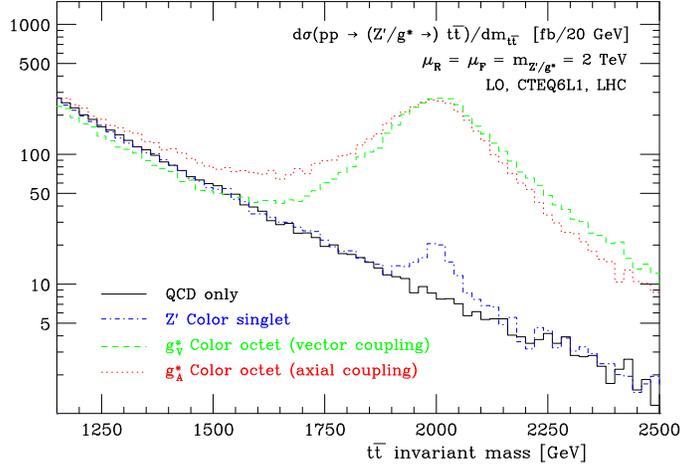}}
  \caption{Top pair invariant mass distribution for color singlet vector and
color octet vector and axial vector resonances \cite{Frederix:2007gi}.}\label{Fig:mttbar}
\end{figure}

Another important distribution for both the Tevatron and the LHC is
in the invariant mass $M_{t\bar{t}}$. The Standard Model shape
has relatively small uncertainty but is sensitive to the top mass,
and may thus assist in determining it.
Shape deviations from the QCD predictions in this distribution 
(peaks, peak-dip structures) are telltales of new physics, such as resonances with various 
spin, parity and color quantum numbers. A study employing
the flexibility of MadGraph 
in a bottom-up approach was performed in Ref.~\cite{Frederix:2007gi},
in which only the most generic aspects of new models are used. 
Fig.~\ref{Fig:mttbar} \cite{Frederix:2007gi} contains the invariant mass
distribution for a number of $s$-channel resonances, showing indeed marked
dependence on the quantum numbers of the resonance.

\subsection{Top spin}
\label{sec:top-spin}

Part of the attractiveness of the top quark as a study object is
its power to self-analyze its spin, through its purely left-handed
SM weak decay. This is both a useful aid in signal-background separations, 
and itself a property worthy of detailed scrutiny, as certain new physics models
could introduce right-handed parts. The correlation between top spin and
directional emission probability for its decay products is expressed
through
\begin{equation}
  \label{eq:13}
  \frac{d\ln \Gamma_f}{d\cos\chi_f} = \frac{1}{2}\left(1+\alpha_f \cos\chi_f  \right)
\end{equation}
where $|\alpha_f| \leq 1$, with 1 indication 100\% correlation.
The dominant decay mode
\begin{equation}
  \label{eq:14}
  t \rightarrow b + W^+ (\rightarrow l^+ + \nu)
\end{equation}
at lowest order, we have $c_b = -0.4, c_\nu=-0.3, c_W = 0.4, c_l=1$. 
QCD corrections to these values are small. The charged lepton direction (or the down-type
quark in a hadronic decay of the intermediate $W$) is indeed 100\% correlated with the
top quark spin. This is amusingly more than for its parent $W$ boson, a consequence of interference
of two amplitudes with different $W$ polarizations. 

\begin{figure}
  \centerline{\includegraphics[width=0.40\columnwidth]{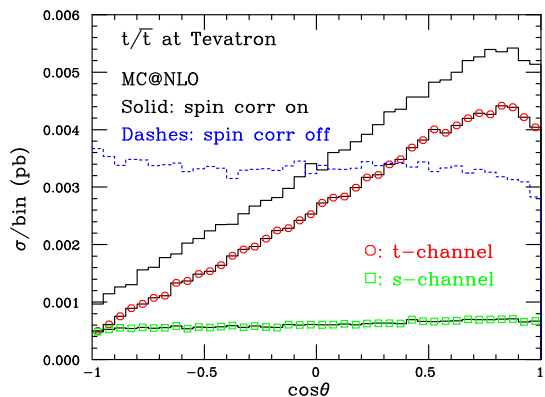}}
  \caption{In $t$-channel single-top production at the Tevatron, a clear correlation
of the lepton flight direction with the recoiling light quark jet. The
correlation disappears when spin-correlations are turned off in \textsc{MC@NLO} \cite{Frixione:2007zp}.} \label{fig:stspincorr}
\end{figure}

In single-top quark production, which occurs via the charged weak
interaction, the top is produced left-handed, so 
a correlation should be a clear feature of the production process
and a discriminant from the background. In Fig.~\ref{fig:stspincorr}
this correlation as computed with \textsc{MC@NLO} \cite{Frixione:2007zp} is shown.
In top quark pair production
a correlation of an individual quark with a fixed direction 
is almost absent , however there is a clear correlation between the top
and anti-top spins. The size of the correlation depends on the choice
of reference axes $\hat{\mathbf{a}},\hat{\mathbf{b}}$ \cite{Bernreuther:2000yn,Bernreuther:2001bx,Mahlon:1997uc}.
At the Tevatron the
beam direction $\hat{\mathbf{a}}=\hat{\mathbf{b}}=\hat{\mathbf{p}}$ is 
good choice, at the LHC the helicity axes 
$\hat{\mathbf{a}}=\hat{\mathbf{b}}=\mathbf{\hat{k}_{top}}$ should give
near-maximal correlation 
\begin{equation}
  \label{eq:12}
  \frac{d\sigma}{d\cos\theta_a d\cos\theta_b}
 = \frac{\sigma}{4}\left(1 + B_1 \cos \theta_a + B_2 \cos \theta_b
 - C \cos\theta_a \cos\theta_b
 \right)
\end{equation}
Indeed, the correlation coefficient $C$ depends in fact on the correlation
axis. Thus, at LO in QCD, the values for 
$\{C_{hel},C_{beam}\}$ at the Tevatron (LHC) is 
$\{=0.47,0.93 \}$ ($\{0.32,-0.01\}$). NLO corrections modify these
number somewhat \cite{Bernreuther:2004jv}. BSM models
that influence the pair production mechanism (e.g. new resonances)
could noticeably influence these correlations.

\section{ASSOCIATED TOP PRODUCTION}
\label{sec:assoc-top-prod}

Many interesting top producing reactions produce other particles in association.
These reactions allow new tests of the top SM interactions,
such as its coupling to the photon, $Z$ or Higgs boson.

Among the most interesting is $pp \rightarrow t\bar{t}H + X$, 
which, if a good sample can be isolated, would allow a direct determination
of the top Yukawa coupling (the SM value is very close to 1). While
it may take some time to gather sufficient data that allow the 
Higgs to be cleanly identified and reconstructed (via
the $H\rightarrow\gamma\gamma$ decay mode) and
backgrounds may be large. A NLO calculation has been carried out
\cite{Beenakker:2002nc,Dawson:2002tg} using a variety of methods,
the $2\rightarrow 3$ kinematics with different masses of the final
state particles making the calculations challenging. 

Study of associated production with an electroweak boson
could reveal anomalous couplings with the top, different from
those in section  \ref{sec:top-quark-standard}. Robust
theoretical tools exist \cite{Baur:2005wi,Baur:2004uw,Lazopoulos:2008de}
which will allow fairly accurate determinations of these 
couplings using LHC data.

Production of $t\bar{t}$ with a jet is another interesting reaction.
Correlations of the top quarks with the extra jet in the final state
in this process can yield a more detailed understanding the top quark
reactions. As such it is a also an important background to BSM processes.
A NLO calculation was completed recently \cite{Dittmaier:2007wz}.
Its role in the NNLO inclusive cross section
calculation  and the determination of the charge asymmetry have been mentioned
already. It is interesting to note that in order to prevent errors in the face of severe
technical complexities, the calculation was purposely carried out, within one collaboration,
in two, fully independent calculational chains .
 
\section{TOP AND MONTE CARLO}
\label{sec:top-monte-carlo}

Perhaps the most widely useful  progress in describing top quark
processes at hadron colliders is in the realm of Monte Carlo. Efforts 
in recent years have led to descriptions beyond 
$2 \rightarrow 2$ processes in 
LO QCD (with subsequent decay and parton showering) in general purpose 
Monte Carlo programs. These fall short when extra hard jets are
present besides the top quarks, nor are they intrinsically normalized
as their only scale dependence in the coupling, with no compensating
terms in the matrix element. Much ingenuity and labor has been brought to
bear to remedy these deficiences. Let us review some of this.

\subsection{Matching}
\label{sec:matching}

Higher-multiplicity  matrix element Monte Carlo's now reach 
$t\bar{t}$ plus up to six jets, and use a variety of methods.
\textsc{ALPGEN} ($t\bar{t}+\leq 6$ jets) does not use Feynman diagrams
but recursion relations to compute the matrix element.
\textsc{COMPHEP} ($t\bar{t}+\leq 1$ jets)  uses squared amplitudes.
\textsc{MADGRAPH/MADEVENT} ($t\bar{t}+\leq 3$ jets) uses complex helicity
amplitudes. However, while  matrix element Monte Carlo's improve the description of 
radiative hard emission emission events, they should if possible not
sacrifice the power of the parton showers to account for collinear and soft
radiation. Matching procedures have been defined to this end.
CKKW \cite{Catani:2001cc} uses $k_T$ clustering to separate
phase space into two regions in each of which one of the descriptions
should hold. To match properly, the matrix elements are reweighted by
Sudakov form factors and $\alpha_s$ factors at the scales correspond
to the nodal branchings.  On the PS side, the showers are vetoed to
ensure that only emissions below the matching scale are included. MLM
\cite{mlm} also reweights the matrix elements, then showers them, but
discards events where the shower generates emission harder than the
matching scale.  Both procedures have been implemented in a number of
matrix-element event generators, and extensively compared \cite{Alwall:2007fs}.

\begin{figure}
  \centerline{\includegraphics[width=0.40\columnwidth]{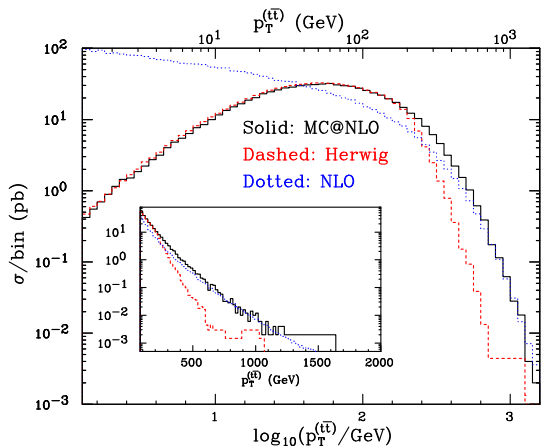}}
  \caption{Top pair transverse momentum distribution at the LHC, showing
agreement with NLO at large, and with parton shower prediction
at small $p_{t\bar{t}}$  \cite{Frixione:2003ei}.}\label{Fig:pttpair}
\end{figure}

\begin{figure}
  \centerline{\includegraphics[width=0.40\columnwidth]{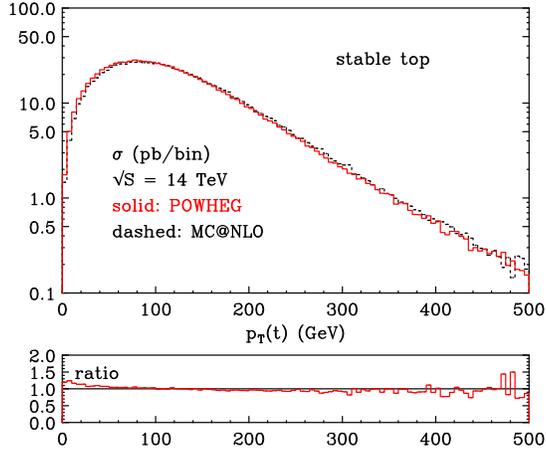}}
  \caption{LHC transverse momentum distribution of top quarks according
to \textsc{MC@NLO} and \textsc{POWHEG}\cite{Frixione:2007nw}.}\label{Fig:psnlotoppt}
\end{figure}

Other very important progress has been in made in matching NLO
to parton shower-based Monte Carlo (\textsc{MC@NLO} \cite{Frixione:2002ik} and \textsc{POWHEG}
\cite{Nason:2004rx}).  Matching is essentially an issue of avoiding
double counting in the one-emission contribution, which can either
come from NLO or from the PS, and in the virtual parts, between the
virtual NLO part and the Sudakov form factors. \textsc{MC@NLO} matches, in
practice, to HERWIG angular-ordered showers.  The benefit
of this matching is clearly visible in Fig.~\ref{Fig:pttpair} \cite{Frixione:2003ei}.
A small percentage of the events \textsc{MC@NLO} generates have a negative weight,
reflecting virtual contributions and subtractions present in NLO and matching.  
\textsc{POWHEG} insists on having positive weights, and exponentiates the complete first order
real matrix element to that end. Both these frameworks are growing in
the list of processes, and realism (e.g. spin correlations
\cite{Frixione:2007zp}). Agreement is generally very good, see Fig.~\ref{Fig:psnlotoppt}, 
also with PS-matched matrix-element generators \cite{Mangano:2006rw},
although interesting differences exist. Such differences reflect genuine ambiguities.

\subsection{$Wt$ production}
\label{sec:wt-production}

\begin{figure}[t]
\centering
\includegraphics[width=0.40\columnwidth]{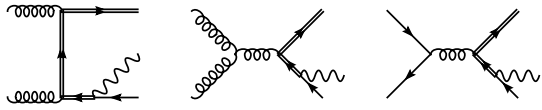}
\caption{Doubly resonant diagrams in NLO corrections to $Wt$ production.}\label{fig:wtlo} 
\end{figure}

An interesting issue arises in the $Wt$ mode of single top production. 
Some diagrams occurring
at NLO contain an intermediate anti-top that can become resonant.
These diagrams can be interpreted as LO $t\bar{t}$ ``doubly resonant'' 
production, with subsequent $\bar{t}$ decay, see Fig.~\ref{fig:wtlo}.
It thus becomes an issue to what extent the $Wt$ and
$t\bar{t}$ can be properly defined as individual processes. 
Several definitions of the $Wt$ channel have been given in the literature,
each with the aim of recovering a well-behaved expansion in $\alpha_s$. The problem
of interference in fact affects any computation that considers contributions
beyond the leading order, i.e. at least ${\cal O}(g_w^2\alpha_s^2)$. The
cross section at this order has been previously presented in
refs.~\cite{Tait:1999cf,Belyaev:1998dn,Kersevan:2006fq}, where only
tree-level graphs were considered, and in 
refs.~\cite{Zhu:2002uj,Campbell:2005bb,Cao:2008af}, where one-loop
contributions were included as well.

\begin{figure}[t]
\centering
\includegraphics[width=0.45\columnwidth]{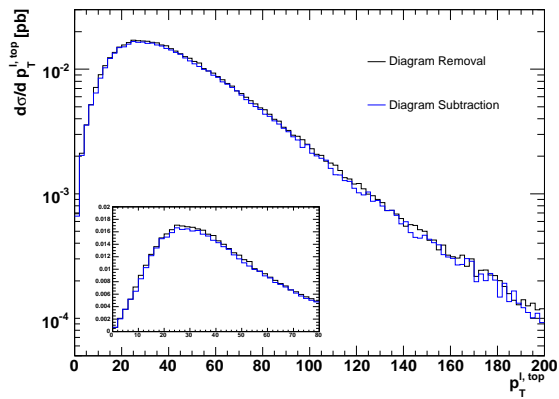}
\caption{$p_T$ spectra in DR and DS scheme for lepton 
from top decay \cite{Frixione:2008yi}.}\label{fig:ptdrds} 
\end{figure}

In Ref.~\cite{Frixione:2008yi}  the issue of interference was addressed
extensively in the context of event generation, in particular the \textsc{MC@NLO} framework. 
Two different procedures for subtracting the doubly-resonant contributions
and recovering a perturbatively well-behaved $Wt$ cross section
were defined. In ``Diagram Removal (DR)''  the graphs in 
Fig.~\ref{fig:wtlo} were eliminated from the calculation, while in ``Diagram Subtraction (DS)''
the doubly resonant contribution was removed via a counterterm. While the former
method is strictly speaking not gauge-invariant, it was shown, first, that gauge variations
are very small, and second that the answer is very close to the gauge invariant
second procedure. The DS procedure leads to the following 
expression for the cross section
\begin{equation}
  \label{eq:5}
  d\sigma^{(2)} + \sum_{\alpha\beta} \int \frac{dx_1dx_2}{x_1x_2S}
\mathcal{L}_{\alpha\beta} \left(\hat{S}_{\alpha\beta} + I_{\alpha\beta}
+ D_{\alpha\beta}-\tilde{D}_{\alpha\beta}  \right) d\phi_3,
\end{equation}
where $\alpha\beta$ labels the initial state channel in which
the doubly-resonant contribution occurs: $gg \; \mathrm{or \; q\bar{q}}$.
$\hat{S}$ is the square of the non-resonant diagrams, 
$I$ their interference with the square of 
graphs of Fig.~\ref{fig:wtlo} $D$. The subtraction term $\tilde{D}$ requires
careful construction.
\begin{equation}
  \label{eq:6}
  \tilde{D}_{gg} = 
\frac{BW (M_{\bar{b}W})}{BW(M_t)} |A_{gg}^{t\bar{t}}|^2_{\mathrm{reshuffled}}
\end{equation}
It was shown that, with suitable cuts, the interference terms are small,
as shown in Fig.~\ref{fig:ptdrds}.
From Eq.~(\ref{eq:5}) it is straightforward to see that the
difference of DR and DS is essentially the interference term.
A particularly suitable cut is a puttting a maximum on
the $p_T$ of the second hardest $b$-flavored hadron, 
a generalization of a proposal made in Ref.~\cite{Campbell:2005bb}.
Thus defined, the $Wt$ and $t\bar{t}$ cross sections
can be separatedly considered to NLO. Their separation at LHC
does remain difficult however.

\section{CONCLUSIONS}
\label{sec:conclusions}

Top quark physics is at present at a pivotal point, 
between the Tevatron and the LHC. As an enterprise
it must progress from promising to performing. The Standard
Model behavior of the top quark has withstood the first 
tests at the Tevatron, and must in the next few years face
a barrage of highly detailed and varied tests by the LHC
experiments. Top's attractiveness as a study object has by 
no means diminished. On the contrary, new observables
are being enlisted to this end. The characteristics
of production and decay, in association with other
particles, are very revealing. Top does not hide
its spin, and awareness of the importance of studying angular distributions
of its decays has grown. 

The theoretical tools for top physics studies are good, 
and keep improving with remarkable pace. With these, and the
turn-on of the LHC at the moment of writing, it is a very safe bet indeed 
to state that the top quark will remain special for years to come.

\begin{acknowledgments}
I would like to thank the organizers for a wonderful and stimulating
conference. I am grateful for clarifying discussions with 
Stefano Frixione, Carlo Oleari and Chris White.
This work has been supported by the Foundation 
for Fundamental Research of Matter (FOM), and by the National Organization 
for Scientific Research (NWO)
\end{acknowledgments}

%% \bibliographystyle{./h-physrev4}
%% \bibliography{./laenen-top.bib}

\end{document}